\documentclass[12pt]{article}
\usepackage{amssymb,latexsym,amsmath}     
\usepackage{authblk}
\usepackage[utf8]{inputenc}
\usepackage{dsfont}
\usepackage[utf8]{inputenc}
\usepackage{graphicx}
\usepackage{mathtools}
\usepackage{amsmath}
\usepackage{subfig}
\usepackage{xcolor}
\usepackage{epstopdf}
\usepackage{placeins}
\usepackage{empheq}
\usepackage{tikz}
\usepackage{siunitx,array,multirow}
\usepackage[colorlinks = true, linkcolor = blue, urlcolor  = blue, citecolor = blue,anchorcolor = blue]{hyperref}
\usepackage[numbers,square,sort&compress]{natbib}
\usetikzlibrary{positioning,decorations.pathmorphing}
\usepackage{pifont}
\usepackage{multirow}
\usepackage[width=\textwidth]{caption}
\usepackage[symbol]{footmisc}
\usepackage{hyperref}
\newcommand{\be}{\begin{equation}}
\newcommand{\nn}{\nonumber}
\newcommand{\bea}{\begin{eqnarray}}
\newcommand{\eea}{\end{eqnarray}}
\newcommand{\ba}{\begin{array}}
\newcommand{\ea}{\end{array}}
\newcommand{\ee}{\end{equation}}

\expandafter\ifx\csname mathbbm\endcsname\relax

\else

\fi \textheight 22cm \textwidth 15cm \topmargin 1mm \oddsidemargin
5mm \evensidemargin 5mm

\newcommand{\beas}{\begin{eqnarray*}}
\newcommand{\eeas}{\end{eqnarray*}}
\newcommand{\bes}{\begin{equation*}}
\newcommand{\ees}{\end{equation*}}

\newcommand{\dir}{{\partial}\!\!\! /}

\newcommand{\lf}{\left}
\newcommand{\ri}{\right}
\newcommand{\f}{\frac}
\newcommand{\dagg}{\dagger}

\def\tr           {\mbox{\rm tr}\,}
\def\det           {\mbox{\rm det}\,}

\def\i2           {\mbox{$\frac{i}{2}$}}

\def\al           {\alpha}

\def\bet           {\beta}

\def\ep           {\epsilon}

\def\et           {\eta}

\def\etb           {{\bar \eta}}

\def\ga           {\gamma}

\def\Ga           {\Gamma}

\def\la           {\lambda}

\def\vph           {\varphi}
\def\ps           {\psi}
\def\psb          {\bar {\psi}}

\def\si           {\sigma}

\def\pl           {\partial}

\begin{document}

\begin{titlepage}
\hfill \vbox{
    \halign{#\hfil         \cr
           } 
      }  
\vspace*{20mm}
\begin{center}
{\LARGE \bf{{On Instantons in Gross-Neveu and \\
\vspace{1.5 mm}
Gross-Neveu-Yukawa models }}}

\vspace*{15mm} \vspace*{1mm} {A. Imaanpur\footnote{aimaanpu@modares.ac.ir} and S. E. Sadati\footnote{e.sadati@modares.ac.ir}}

\vspace*{1cm}

{\it Department of Physics, School of Sciences\\ 
Tarbiat Modares University, P.O. Box 14155-4838, Tehran, Iran}\\
\vspace*{1mm}

\vspace*{1cm}

\end{center}

\begin{abstract}
We study fermionic instantons of the Gross-Neveu and the Gross-Neveu-Yukawa models. We derive solutions for both models and examine the corresponding actions at the fixed points. We further map the solutions on to the sphere and discuss the relation to the Hubbard-Stratonovich approach. Close to the fixed points we compare and identify the results with those obtained in the large $N$ computation. 
\end{abstract}

\end{titlepage}

\section{Introduction}
$O(N)$ vector models have served many applications in particle physics as well as in condensed matter physics. They provide a natural framework to study spontaneous symmetry breaking and critical phenomena. Moreover, they have also appeared in AdS/CFT duality as theories that describe the boundary physics \cite{KLPO}. Recently, it has been observed that $O(N)$ models show a range of interesting low energy features in the space dimensions $4\leq d \leq 6$ \cite{KLON}. The corresponding instantons and their nonperturbative effects have further been discovered and compared with the large $N$ computations \cite{KLCO}. 

In the Gross-Neveu model fermionic fields take over the bosonic vector fields of $O(N)$ models \cite{GRNE}. This model has been used as a toy model to study chiral symmetry breaking and the strong coupling phenomena in the low energy limit \cite{ZJFO, MOSH}. In this article we study instanton solutions of the Gross-Neveu and the Gross-Neveu-Yukawa models. These models are renormalizable in $d=2$ and $d=4$, respectively, and show stable fixed points. As fermions are the basic degrees of freedom, the solutions we find are fermionic. However, there is a peculiarity when one studies fermions in the Euclidean signature; the action is not Hermitian under the sole Hermitian conjugation of the fields. Hermiticity is achieved by a further space inversion along one of the space directions. This symmetry is referred to as {\em reflection Hermiticity} \cite{ZJB}. We will see that the equations of motion of $\psi$ and $\psb$ inherit the symmetry, and because of that, solutions do not exist for which $\psb\psi$ is real. To get around this problem, we consider $\psi$ and $\psb$ to be independent fields, and then, to avoid doubling the degrees of freedom, we may further impose the Majorana or Weyl conditions. In fact, this is customary in studying quantum field theories in the Euclidean signature. For a complex scalar field $\vph$, for instance, one could either take the real and imaginary parts of $\vph$ as independent fields, or, consider $\vph$ and $\bar{\vph}$ to be independent degrees of freedom. By doing so, we find consistent smooth solutions with finite actions. Since the equations of motion are conformally invariant, we can map the solutions on to  $S^d$ obtained from $E^d$ by a stereographic projection. As in the case of instantons in $O(N)$ models \cite{KLCO}, we observe that for a specific choice of the moduli the solutions on the sphere become constant. And, this is an indication that it might be possible to get the same constant solutions in the large $N$ limit of the Hubbard-Stratonovich approach.  

In the next Section, we discuss the Gross-Neveu model in the Euclidean space and its reflection Hermiticity. The symmetry is reflected on the equations of motion, and thus restricts the form of the solutions. We derive spinorial instanton solutions and compute their actions at the UV fixed point. Instantons of the Gross-Neveu-Yukawa are studied next in Section 3. In Section 4, we discuss mapping the solutions on to $S^d$, and observe that for a specific choice of the moduli they turn out to be constant. In addition, we make use of the eigenfunctions of the Dirac operator on $S^d$ to derive more generic solutions of this kind. In Section 5, we use the Hubbard-Stratonovich approach to derive the constant critical points of the effective action in the large $N$ limit. We examine the critical points and observe that they are in close correspondence with the instanton solutions we derived in the previous Sections. Conclusions and out look are brought in Section 6.

\section{Gross-Neveu Model}
In this Section, we start with a discussion of the Gross-Neveu model in the Euclidean space. In particular, we illustrate the concept of reflection Hermiticity and its effect on the solutions of the equations of motion. Next, we derive a set of fermionic instantons with the action being a function of the coupling constant. At the end, to make contact with the Hubbard-Stratonovich approach, the action is computed at the fixed point in $d=2+\ep$. 

The Gross-Neveu model \cite {GRNE} consists of a spinor field with a quartic self-interaction term, which is renormalizable in two dimensions. 
One could further take the spinor field to be an ${\tilde N}$-component vector and thus increase the symmetry of the action to $U({\tilde N})$. The action is as follows
\be
S=\int d^2x \lf[ \ i\psb\cdot \dir \psi +\f{g}{2} (\psb\cdot \psi)^2 \ri] ,\label{GMIN}
\ee
with $\psb =\psi^\dagg \ga_0$. Here $\psi$ is an ${\tilde N}$-component vector, i.e., $\psi=(\psi_1,\ldots,\psi_{\tilde N})$, and we use the notation $\psb\cdot\psi =\sum_{i=1}^{\tilde N} \psb_i \psi_i$. 

In the Euclidean space, however, the action is taken to be
\be
S=-\int d^2x \lf[ \ \psb\cdot \dir \psi +\f{g}{2} (\psb\cdot \psi)^2 \ri] ,\label{ACT}
\ee
where the Gamma matrices are Hermitian, and $\psb$ is defined by the Hermitian conjugation followed by a space inversion along one of the coordinates, say $x_d$, (active point of view) through \cite{ZJB},
\be
\psi(x)\to \ga_d \psi(\tilde{x})\, ,\ \ \ \ \psb (x) \to \psb (\tilde{x}) \gamma_d\, ,\label{INV}
\ee
here $\tilde{x}=(x_1,\ldots,- x_d)$ is the space inversion of $x=(x_1,\ldots,x_d)$ along the $d$ direction.

Let $\mu=1,\ldots,d$, and $m=1,\ldots, d-1$. We observe that under the Hermitian conjugation the kinetic term transforms as
\be
\psb\cdot \ga^\mu \pl_\mu \psi (x) \to \pl_\mu\psb\cdot \ga^\mu \psi(x)\, .
\ee
Next, doing the space inversion (\ref{INV}), yields
\bea
\pl_\mu\psb\cdot  \ga^\mu \psi(x)&\rightarrow & (\pl_m\psb (\tilde{x}) \ga_d \ga^m +\pl_d \psb (\tilde{x}) \ga_d\ga^d)\cdot \ga_d\psi (\tilde{x}) \nn \\
\!\!&=&\!\!(-\pl_m\psb (\tilde{x})\ga^m +\pl_d\psb(\tilde{x})\ga^d)\cdot \psi (\tilde{x})\nn \\
\!\!&=&\!\!(-\pl_m\psb (\tilde{x})\ga^m -\tilde{\pl}_d\psb(\tilde{x})\ga^d)\cdot \psi (\tilde{x})\nn \\
\!\!&=&\!\! -\pl_\mu\psb\cdot\ga^\mu \psi (\tilde{x})\, ,
\eea
where in the second line we used $\{\ga_m , \ga_d\}=0$, and $\ga_d^2=1$. Transferring the partial derivative to $\psi$ under the integral sign with a change of coordinates $(x_1,\ldots,-x_d)\to (x_1,\ldots, x_d)$ results in
\be
\int d^2x \ \psb\cdot \dir \ps \to \int d^2x \ \psb\cdot \dir \ps\, .
\ee
The interaction term under Hermitian conjugate transforms as
\be
\psb\cdot \psi (x)\rightarrow \psb\cdot\psi(\tilde{x})\, ,
\ee
so that its integral is invariant. Hence, we conclude that action (\ref{ACT}) is symmetric under the Hermitian conjugation followed by space inversion. This symmetry is referred to as  {\em reflection Hermiticity}.

\subsection{Instanton solutions}
Let us begin our discussion by the analysis of the equations of motion. They read, for $\psb$ and $\psi$, respectively,
\bea
\ga^\mu \pl_\mu \psi (x) = -g (\psb\cdot\psi)\, \psi (x)\, \label{eq1}\, ,\\
\pl_\mu \psb (x) \ga^\mu = g (\psb\cdot\psi)\, \psb (x) \, .\label{eq2}
\eea
Since the action has the reflection Hermiticity symmetry, it can be seen that the above equations are also related by this symmetry. In particular, note that the second equation is {\em not} the Hermitian conjugate of the first (in the Lorentzian signature, with  an $i$ factor included, and $\psb=\psi^\dagg \ga_0$, the two equations are related by the Hermitian conjugation). This implies that in looking for solutions one cannot simply take $\psb=\psi^\dagg$, as the above equations would require $\psi^\dagg\psi$ to be pure imaginary. Therefore, to find a consistent solution, we take $\psi$ and $\psb$ to be two independent spinors. 

To proceed, we make the following ansatz
\be
\psi= \f{C+i\si\cdot (x-x_0)}{C^2+(x-x_0)^2}\, \xi\, ,\ \ \ \ \ \ \psb =\etb\, \f{C-i\si\cdot (x-x_0)}{C^2+(x-x_0)^2}\, ,\label{ANS}
\ee
where $C$ is a real constant, $\si^\mu =(\si^1, \si^2)$ are the first two Pauli matrices, and, $\xi$ and $\eta$ are two constant ${\tilde N}$-component spinors.\footnote{ Let $
\xi=\lf(
\begin{array}  {l}
\al \\ 
\bet 
\end{array}
\ri) .
$ To keep the degrees of freedom unchanged, one might, for instance, take $\et=\si_2 \xi$. Then equation (\ref{C}) fixes the imaginary part of $\al\bet^*$.} Using $(\si\cdot x)^2=x^2=x_1^2+x_2^2$, we derive
\be
\dir \psi =\f{2iC(C+i\si\cdot (x-x_0))}{(C^2+(x-x_0)^2)^2}\, \xi\, ,\label{DIR}
\ee
and,
\be
\psb\cdot \psi= \f{1}{C^2+(x-x_0)^2}\, \etb\cdot \xi\, .\label{NORM}
\ee
Plugging (\ref{DIR}) and (\ref{NORM}) into (\ref{eq1}) implies
\be
\etb\cdot \xi=-\f{2iC}{g}\, .\label{C}
\ee
The same result follows from (\ref{eq2}). Had we taken $\psi$ and $\psb$ to be complex conjugate of each other the above equation would have been inconsistent and thus no solution (taking $C$ to be imaginary would result in a singular solution). 

Let us then compute the action of such instantons. Using the equation of motion (\ref{eq1}), together with (\ref{NORM}) and (\ref{C}), the on-shell action becomes
\bea
S&=&-\int d^2x \lf[ \ \psb\cdot \dir \psi +\f{g}{2} (\psb\cdot \psi)^2 \ri]=+\f{g}{2}\int d^2x\, (\psb\cdot \psi)^2 \nn \\
&=&-\f{2C^2}{g}\int \f{d^2x}{(C^2+(x-x_0)^2)^2}=-\f{2\pi}{g}\, .\label{S}
\eea
Note that $S$ does not depend on $C$ or $x^\mu_0$. This shows that $C$ and $x^\mu_0$ are the moduli of the instanton solution. Furthermore, we also have two spinor moduli $\xi$ and $\eta$ which are constrained and related to the moduli $C$ through (\ref{C}).

As mentioned in the beginning, we will need the value of the action at the fixed point of the model to compare with the Hubbard-Stratonovich approach in Section 5. In $d=2+\ep$, the Gross-Neveu model has a UV fixed point \cite{MOSH};
\be
g^* =\f{2\pi \ep}{N}\, , \label{FIX1}
\ee
note that here $N=2 {\tilde N}$, where the extra $2$ accounts for the dimension of spinors. In a general dimension $d$, we have $N=2^{[d/2]}{\tilde N}$. 
At this fixed point, action (\ref{S}) becomes
\be
S=-\f{N}{\ep}\, .\label{S1}
\ee
We will return to this result in Section 5. 

\section{Gross-Neveu-Yukawa Model}
The Gross-Neveu-Yukawa Model is characterised by an extra scalar field $\si$ which is coupled to fermions through a Yukawa coupling. This model shows an infrared fixed point in $d=4-\ep$, and thus, is believed to be the UV completion of the Gross-Neveu model. The action reads
\be
S=\int d^4x \, \lf[-\psb\cdot (\dir + g_1 \si)\ps +\f{1}{2}(\pl_\mu \si)^2 +\f{g_2}{4!}\si^4 \ri] ,
\ee
with the equations of motion
\bea
&&(\dir +g_1 \si)\psi =0\, ,\label{E1} \\
&& \pl^2\si -\f{g_2}{6}\si^3 +g_1 \psb\cdot\psi=0 \, .\label{E2}
\eea

First, we note that there is a {\em bosonic} instanton 
\be
\si=\sqrt{\f{48}{-g_2}} \f{C}{1+C^2 (x-x_0)^2}\, , \ \ \ \ \ \ \psi =0\, ,\label{BOS}
\ee
with $x^2=\sum_{i=1}^4 x_i^2$. This solution was also studied in \cite{MCK, PET, KLCO}. The on-shell action for this solution is
\be
S=-\f{g_2}{4!}\int d^4x \, \si^4 = -\f{96C^4}{g_2} \int \f{d^4x}{(1+C^2 (x-x_0)^2)^4} = -\f{16\pi^2}{g_2}\, .
\ee
Taking $d=4-\ep$, the renormalization group equation has an infrared fixed point \cite{MOSH};
\be
(g_1^*)^2=\f{16 \pi^2 \ep}{N}\, , \ \ \ \ \ \ g_2^* =16 \pi^2 R \ep\, ,\label{FIX}
\ee
where,
\be
R=\f{24 N}{(N+6)[(N-6)\sqrt{N^2+132N +36}]}\, .
\ee
Note that here $N$ is related to ${\tilde N}$ through $N=4{\tilde N}$.
In the large $N$ limit
\be 
R\sim \f{12}{N}\, ,
\ee
and so,
\be
g_2^* =192 \pi^2 \f{\ep}{N}\, .
\ee
Hence the action at the fixed point becomes
\be
S=-\f{N}{12\ep}\, .\label{S2}
\ee

Turning on the spinors, we make a similar ansatz to the one in the previous Section
\be
\psi= \f{C+i\si\cdot (x-x_0)}{(C^2+(x-x_0)^2)^2}\, \xi\, ,\ \ \ \ \psb =\etb\, \f{C-i\si\cdot (x-x_0)}{(C^2+(x-x_0)^2)^2}\, , \ \ \ \ \si = \f{B}{C^2+(x-x_0)^2}\, ,\label{ANS2}
\ee
where $B$ is a new constant. Using this ansatz in (\ref{E2}), we get
\be
8B C^2 +\f{g_2}{6}B^3 -g_1 \etb\cdot \xi=0\, ,\label{BC}
\ee
whereas, (\ref{E1}) requires
\be
B=-\f{4iC}{g_1}\, .\label{Cg}
\ee
Plugging (\ref{Cg}) into (\ref{BC}) gives 
\be
\etb\cdot\xi=-32i C^3\lf(\f{1}{g_1^2}-\f{g_2}{3 g_1^4}\ri) .
\ee
The on-shell action for this solution is calculated to be
\bea
S&=&\int d^4x \, \lf[-\f{1}{2}\si\pl^2 \si +\f{g_2}{4!}\si^4 \ri]=\int d^4x \, \lf[\f{1}{2} g_1\si\psb\cdot\psi -\f{g_2}{4!}\si^4 \ri] \nn \\
&=& 32C^4\lf(\f{g_2}{3g_1^4}-\f{2}{g_1^2}\ri) \int \f{d^4x}{(C^2+(x-x_0)^2)^4} \nn \\
&=& \f{32 \pi^2}{3}\lf(\f{g_2}{6g_1^4}-\f{1}{g_1^2}\ri).
\eea
At the fixed point, (\ref{FIX}), and the large $N$ limit, the action reads
\be
S= \f{2N}{3\ep}\, .\label{S3}
\ee
We postpone a further discussion of the results of this Section to Section 5, after introducing the Hubbard-Stratonovich method.

\section{Mapping on to $S^d$}
We could map the solutions we derived in the previous Sections on to the sphere, $S^d$. The reason for this mapping, as we will see, is that on a particular point of the moduli space the solutions on the sphere turn out to be constant, and, this is a hint that one might be able to get the same constant solutions as the critical points in the Hubbard-Stratonovich approach \cite{KLCO}. To start with, note that the metric of $S^d$ can be obtained by a stereographic projection of $d$-dimensional Euclidean flat space
\be
ds^2_{S^d}= \f{4}{(1+x^2)^2}\sum_{i=1}^d dx_i^2\, .
\ee
Therefore, since coordinates scale by a factor of $\f{2}{1+x^2}$, fields with a mass dimension $\Delta$ transform as
\be
\vph \to \lf(\f{1+x^2}{2}\ri)^\Delta \vph \, .
\ee
Moreover, since the equations of motion are conformally invariant, this maps solutions on flat space, $E^d$, to solutions on $S^d$.

In $d=2$, spinors have dimension $1/2$ so that $\psb\psi$ in (\ref{NORM}) under a Weyl rescaling to $S^2$ transforms as
\be
\psb\cdot \psi= \f{1}{C^2+(x-x_0)^2}\, \etb\cdot \xi\, \to  \f{1+x^2}{2(C^2+(x-x_0)^2)}\, \etb\cdot \xi\, ,
\ee
hence, for the special choice of $C=1$ and $x_0=0$, it becomes a constant
\be
\psb\cdot \psi= \f{1}{2}\, \etb\cdot \xi\, ,
\ee
with the same action as in (\ref{S1}), because of the conformal invariance. 

In $d=4$, the dimensions of $\si$ and $\psi$ are $1$ and $3/2$, respectively. So, the bosonic instanton, (\ref{BOS}), on $S^4$ is obtained by the following scaling
\be
\si\to \f{1}{2}\sqrt{\f{48}{-g_2}} \f{C(1+x^2)}{1+C^2 (x-x_0)^2}\to \si =\f{1}{2}\sqrt{\f{48}{-g_2}} \, ,
\ee
where, in the last step we have set $C=1$ and $x_0=0$. Similarly, the spinorial instanton, (\ref{ANS2}), scales to
\be
\si\to \f{B}{2}\, , \ \ \ \ \psb\psi \to \f{\etb\xi}{8}\, ,
\ee
and, of course, with the same actions we derived in (\ref{S2}) and (\ref{S3}). The upshot is that for a specific choice of the moduli, i.e., $C=1$ and $x_0=0$, the solutions turn out to be constant on the sphere. This analysis shows that constant solutions on the sphere must somehow show up as  saddle points in the path integral formalism. In the next section, we will see that this is the case; the constant solutions appear as the critical points of the effective action in the Hubbard-Stratonovich approach.

\subsection{Solutions by Dirac operator eigenfunctions}
We can derive more solutions by using the eigenfunctions of the Dirac operator on the sphere. Starting with $d=2$, note that equation of motion (\ref{eq1}), when 
$\psb\cdot\psi$ is constant, becomes an eigenvalue equation. However, the eigenvalues and eigenfunctions of the Dirac operator on $S^d$ are known:
\be
i\nabla\!\! \!\!\hspace{-0.2 mm} / \, \psi_{n,p} = \la_n \psi_{n,p}\, , 
\ee
with 
\be
\la_n=\pm(n+\f{d}{2})\, ,\label{LAM}
\ee
and $p=1,\ldots,d_n$ labeling the degeneracies at level $n$
\be
d_n=\f{\Ga(n+d)}{n!\Ga(d)}\, .\label{DEG}
\ee
So, we could arrange the ${\tilde N}$ components of $\psi$ as follows
\be
\psi =(\psi_{n,1},\ldots,\psi_{n,d_n},0,\ldots,0)\, ,\label{SOLEIG}
\ee
assuming ${\tilde N}\geq d_n$. And then, use the addition theorem for $\psi_{n,p}$'s at a fix $n$ to normalize $\psi$ as
\be
i g\, \psb\cdot \psi = (n+1)\, ,
\ee
therefore, $\psi$, as constructed in (\ref{SOLEIG}), is a solution to equation of motion (\ref{eq1}). The on-shell action becomes
\bea
S&=&-\int d^2x\sqrt{g} \lf[ \ \psb\cdot \nabla \!\!\!\!\hspace{-0.2 mm} /\, \psi +\f{g}{2} (\psb\cdot \psi)^2 \ri]=+\f{g}{2}\int d^2x \sqrt{g}\, (\psb\cdot \psi)^2 \nn \\
&=&-\f{2\pi}{g}(n+1)^2\, .
\eea
In $d=2+\ep$, and at the fixed point, (\ref{FIX1}), we have
\be
S=-\f{N}{\ep}(n+1)^2\, .\label{S4}
\ee

In $d=4$, for the Gross-Neveu-Yukawa model on $S^4$, equation of motion (\ref{E1}) becomes an eigenvalue equation for the Dirac operator if we set
\be
ig_1\si = (n+2)\, , 
\ee
and then, equation of motion of $\si$, (\ref{E2}), fixes the normalization of $\psi$. The on-shell action for this constant solution becomes
\bea
S&=&\int d^4x\sqrt{g} \, \lf[\f{1}{2}\pl_\mu\si\pl^\mu \si +\si^2+\f{g_2}{4!}\si^4 \ri] \nn \\
&=&\f{8\pi^2}{3} \, \lf[ -\f{(n+2)^2}{g_1^2}+\f{g_2}{4!}\f{(n+2)^4}{g_1^4}\ri] ,
\eea
where, the second term in the first line is coming from the conformal coupling of scalars on $S^4$. At the fixed point, (\ref{FIX}), and the large $N$ limit, the action reads
\be
S= -\f{N(n+2)^2}{6\, \ep}\lf(1-\f{(n+2)^2}{2}\ri) .\label{S5}
\ee
Notice that for $n=0$ the above actions, (\ref{S4}) and (\ref{S5}), coincide with actions (\ref{S1}) and (\ref{S3}) we derived for instantons on flat space. 

\section{Hubbard-Stratonovich approach}
Hubbard-Stratonovich approach is an effective description of the Gross-Neveu model which allows to study it in the continuous range of $2\leq d \leq 4$, and, also examine its relation to the Gross-Neveu-Yukawa model. In particular, if we could derive an instanton solution in this approach it exists for the whole range of $d$. Further, since, in general, the action of instantons is a function of the coupling constants, this approach gives valuable information on the existence and location of the fixed points. This approach has been widely used in \cite{KLCO, KLIN}  to discuss the interrelations of the $O(N)$ models and their instanton solutions in $4\leq d \leq 6$. Here we adopt a similar method.    

Let us first consider once again the Gross-Neveu model, but this time in a general dimension $d$,
\be
S=-\int d^dx \lf[ \ \psb\cdot \dir \psi +\f{g}{2} (\psb\cdot \psi)^2 \ri] .\label{ACT2}
\ee
We could use an auxiliary scalar field $\si$, the Hubbard-Stratonovich field, to rewrite the above action as 
\be
S=\int d^dx \, \lf[-\psb\cdot (\dir +  \si)\psi +\f{\si^2}{2g} \ri] .\label{HS}
\ee
By using the equation of motion of $\si$ and then replacing $\si$ back into the action we get back to (\ref{ACT2}), therefore, the two actions are equivalent. However, 
action (\ref{HS}) is quadratic in the fermionic fields, and hence, the gaussian path integral could be performed over them to get an effective action for $\si$:
\be
{\cal S}= -{\tilde N}\log \det (\dir +\si) + \f{\si^2}{2g}\, .
\ee
The second term will be irrelevant at the infrared fixed point \cite{KLCO} and we ignore it in the following discussion. Let us then define
 \be
{\cal S}= -{\tilde N}\log \det (\dir +\si) = N F_f(\si)\, ,\label{F-f}
\ee
where $F_f(\si)$ stands for the free energy of a single spinorial component of $\psi_i$, and the subscript $f$ for fermions. Now, we see that at the large $N$ limit, the path integral over $\si$ localizes on the critical points of $F_f(\si)$.

On $S^d$, we can use the eigenvalues and degeneracies of the Dirac operator, (\ref{LAM}) and (\ref{DEG}), to compute $F_f(\si)$ for a constant $\si$
\bea
F_f(\si)&=&-\f{1}{\tr {\bf 1}}\log \det (\nabla\!\! \!\!\hspace{-0.2 mm} /  +\si) \\
&=&- \log \prod_{n=0}^\infty ((i(n+d/2)+\si)(-i(n+d/2)+\si))^{d_n} \\
&=&-\sum_{n=0}^\infty d_n \lf(\log (n+d/2 +i\si) +\log (n+d/2 -i\si) \ri) ,
\eea
the first line follows because of the definition in (\ref{F-f}), and $N=2^{[d/2]} {\tilde N}=\tr {\bf 1}\,  {\tilde N}$. To find the critical points, let us then take the derivative of $F_f(\si)$ 
\bea
\f{\pl{F}_f(\si)}{\pl \si} &=&-\sum_{n=0}^\infty d_n\lf(\f{i}{  n+d/2 +i\si}-\f{i}{n+d/2-i\si}\ri) \nn \\
&=& -\f{i}{\Ga(d)}\sum_{n=0}^\infty \f{1}{n!}\lf(\f{\Ga(n+d)}{  n+d/2 +i\si}-\f{\Ga(n+d)}{n+d/2-i\si}\ri).\label{F}
\eea
However, using
\be
\sum_{n=0}^\infty \f{\Ga(n+a)}{n!}\f{1}{n+b} = \f{\pi}{\sin \pi a} \f{\Ga(b)}{\Ga(1-a+b)}\, ,
\ee
together with
\be
\Ga(1-z)\Ga(z)=\f{\pi}{\sin \pi z}\, ,
\ee
(\ref{F}) can be simplified. In the following, to keep track of the singularities when $\ep$  in $d$ goes to zero, it is useful to define $\tilde{F}_f(\si)=-\sin( \pi d/2) F_f(\si)$. So, we get
\be
\f{\pl \tilde{F}_f(\si)}{\pl \si} =\f{1}{\Gamma(d)}\Ga\lf(\f{d}{2}+i \si\ri) \Ga\lf(\f{d}{2}-i \si\ri) \sinh \pi \si\, ,\label{GAM}
\ee
which has zeros at
\be
\si=i k\, ,
\ee
for $k$ an integer. Note that when $d$ is close to integers, e.g. $d=2+\ep$, or $d=4-\ep$, depending on the sign of $k$, one of the Gamma functions in (\ref{GAM}) behaves like $1/\ep$ while the other one is finite as $\ep \to 0$. Therefore, the Gamma functions for integers $k$ are finite and the zeros of (\ref{GAM}) happen only when $\sinh \pi \si$ becomes zero.

For integer $d$, eq. (\ref{GAM}) can further be simplified. Using $\Ga(1+z)=z\Ga(z)$, together with
\be
|\Ga(1+i\si)|^2=\f{\pi \si}{\sinh \pi \si}\, ,
\ee
we see that for $d=2$, (\ref{GAM}) becomes
\be
\f{\pl \tilde{F}_f(\si)}{\pl \si} =\pi \si\, .
\ee
Therefore, by integration we arrive at
\be
\tilde{F}_f(\si)-\tilde{F}_f(0)=\f{\pi \si^2}{2}\, .
\ee
For $k=1$, we have $\si^2=-1$, and thus the action, subtracting the contribution of the trivial vacuum, reads
\be
{\cal S}_{\si=i}-{\cal S}_{\si=0}=N(F_f(i)-F_f(0))= \f{2N}{\pi \ep} (\tilde{F}_f(i)-\tilde{F}_f(0)) =-\f{N}{\ep}\, ,
\ee
which is consistent with result (\ref{S1}) we derived in Section 2. For $\si=i k$, we get
\be
{\cal S}_{\si=ik}-{\cal S}_{\si=0} = -\f{N k^2}{\ep}\, .
\ee
Setting $k=n+1$, the above result coincide with (\ref{S4}), which was derived using the eigenfunctions of the Dirac operator on $S^2$.

For $d=4$, (\ref{GAM}) reduces to
\be
\f{\pl \tilde{F}_f(\si)}{\pl \si} =\f{\pi}{6}(\si+\si^3)\, ,
\ee
and hence
\be
\tilde{F}_f(\si)-\tilde{F}_f(0)=\f{\pi }{12}(\si^2+\f{\si^4}{2})\, .
\ee
Here if we set $k=1$, i.e. $\si^2=-1$, we get for the action
\be
{\cal S}_{\si=i}-{\cal S}_{\si=0}=N(F_f(i)-F_f(0))= \f{2N}{\pi \ep} (\tilde{F}_f(i)-\tilde{F}_f(0))=-\f{N}{12\ep}\, ,
\ee
which is the action we derived for bosonic solution in (\ref{S2}). For $k=2$, $\si^2=-4$, and we derive
\be
{\cal S}_{\si=2i}-{\cal S}_{\si=0}=\f{2N}{3\ep}\, ,
\ee
which is the result we derived from spinorial instantons in (\ref{S3}). Note that if we set $\si=ik=i (n+2)$, the action reads 
\bea
{\cal S}_{\si=ik}-{\cal S}_{\si=0}&=&N(F_f(ik)-F_f(0))= \f{2N}{\pi \ep} (\tilde{F}_f(ik)-\tilde{F}_f(0))\nn \\
&=&\f{N }{6 \ep}(-k^2+\f{k^4}{2})=-\f{N(n+2)^2 }{6 \ep}\lf(1-\f{(n+2)^2}{2}\ri) ,
\eea
precisely the result we derived in (\ref{S5}), using the eigenfunctions of the Dirac operator on $S^4$.

\section{Conclusions}
In this paper we examined the existence of fermionic instantons in the Gross-Neveu and the Gross-Neveu-Yukawa models. First, we derived some solutions on the Euclidean flat space with the corresponding moduli. We computed their actions at the fixed points. This step allowed us to compare the result with the Hubbard-Stratonovich approach. To do so, we mapped the solutions on to the sphere and observed that at a specific point in the moduli space the solutions become constant, a hint that they might be observable in the large $N$ limit. Since the Hubbard-Stratonovich method is applicable for a generic spacetime dimension, we could analytically continue those constant solutions to $d=2+\ep$ and $d=4-\ep$, at which we expected the emergence of the fixed points. This enabled us to compare the action of these constant critical points of the effective action with those of the fermionic instantons at such fixed points. Our analysis, in particular, showed the power of Hubbard-Stratonovich method in understanding the nonperturbative effects in the large $N$ limit. The method has so far proved useful to learn about the existence and location of the fixed points. 

The instanton solutions we explored in this paper will be essential to further investigate the nonperturbative effects through a semiclassical expansion around them. Especially, instanton contributions to the anomalous dimensions and the current correlation functions are among the interesting problems to be studied.

\end{document}